\documentclass[intlimits,twoside,a4paper]{article}
\usepackage{amsmath,amssymb}
\usepackage{graphicx}
\usepackage{wrapfig}
\usepackage{bm}
\usepackage[T2A]{fontenc}
\usepackage[cp1251]{inputenc}

\usepackage{threeparttable}

\usepackage[eqsecnum]{cmpj}

\issue{2011}{14}{4}{43301}

\doinumber{10.5488/CMP.14.43301}

\title[Bose-Fermi-Hubard model on two sublattices]%
{Bose-Fermi-Hubbard model on a lattice with two nonequivalent sublattices}

\author[T.S. Mysakovych]{T.S. Mysakovych}

\authorcopyright{T.S. Mysakovych, 2011}

\address{
Institute for Condensed Matter Physics of the National Academy of Sciences of Ukraine,\\
1 Svientsitskii Str., 79011 Lviv, Ukraine
}

\date{Received July 20, 2011, in final form September 14, 2011}

\begin{document}

\maketitle

\begin{abstract}

Phase transitions  in systems described by Bose-Fermi-Hubbard model
on a lattice with two nonequivalent sublattices are investigated in this work.
The case of hard-core bosons is considered and pseudospin formalism is used.
Phase diagrams  are built in the plain of chemical potential of the bosons-bosonic hopping parameter.
It is shown that in the case of anisotropic hopping,
the region of the supersolid phase existence is possible
for a smaller parameter space.
\pacs 37.10.Jk, 67.85.Pq, 71.10.Fd

\end{abstract}

\section{Introduction}

Theoretically, the systems of ultracold atoms are well
described by the Hubbard model. For the case of  bosonic atoms,  the  investigations of the Bose-Hubbard
model  predicted  a superfluid to Mott insulator transition in ultracold
atomic gases~\cite{fisher,krishnamurthy}.
The experimental verification of this
phase transition was done by Greiner et al.~\cite{greiner} and the study of this model
has rapidly developed since then. Besides the bosonic atoms, the mixture of bose-fermi atoms in optical lattices was realized and the Bose-Fermi-Hubbard model (BFHM) is under intensive investigations~\cite{albus,blatter,lewenstein,freericks,titvinidze,mering2,mysakovych1}. As concerns the BFHM,  phase diagrams are more complicated because the effective interaction between bosons is generated due to the presence of fermions and this leads, for example, to the possible appearance of a supersolid (SS) phase (which is characterized by a simultaneous presence of a density wave and phase order in the condensate).

The lattice depth, dimensionality, geometry, and filling factor can all be controlled.
Apart from the  uniform lattice potentials, other lattice topologies can be realized.
For example, the superposition of two standing-wave lattices of different wavelengths leads to a
superlattice with a spatial modulation of the lattice well depths.
In~\cite{sebby}
an optical lattice of double wells  was realized by combining laser beams with the in-plane and out-of-plane polarizations of light.
 The properties of the system of
bosonic atoms in superlattice potential were studied in~\cite{aizeman,buenoz,iskin,rigol,qian,rigol2},
where the presence of the modulated potential leads to the spatial modulation of the bosonic concentration.
In~\cite{roth} the zero temperature phase diagram of binary boson-fermion mixtures
in one-dimensional superlattices was investigated using an exact numerical diagonalization
technique. It should be noted that almost all the mentioned studies were restricted to the case of zero temperature. In our previous work~\cite{mysakovych1} we analyzed phase transitions in the BFHM at finite temperature.

The Bose-Fermi-Hubbard-type model can  also be applied in the description of intercalation of ions in crystals
(for example, lithium intercalation in TiO$_2$ crystals).
Ion-electron interaction can play a significant role in these systems.
At intercalation, the chemical potential in such systems is displaced into the
conduction band. Phase separation into Li-poor  and Li-rich  phases occurs in such crystals and  this two-phase
behaviour leads to a constant value of the electrochemical
potential~\cite{wagem_2001,wagem_2003} (this is taken in consideration when constructing
batteries).

\section{Model and results}

\begin{figure}[ht]
\includegraphics[width=0.45\textwidth]{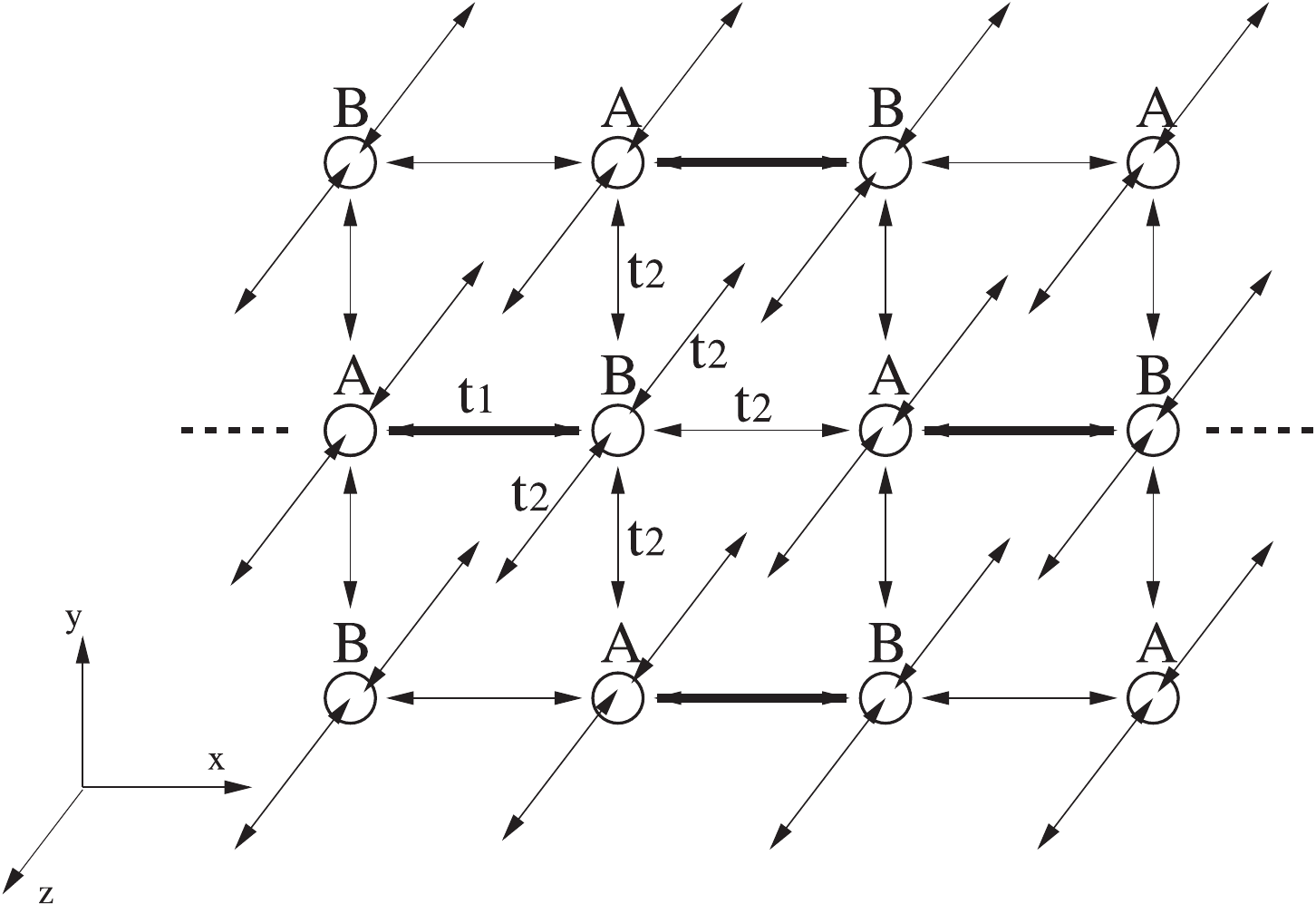}%
\hfill%
\includegraphics[width=0.45\textwidth]{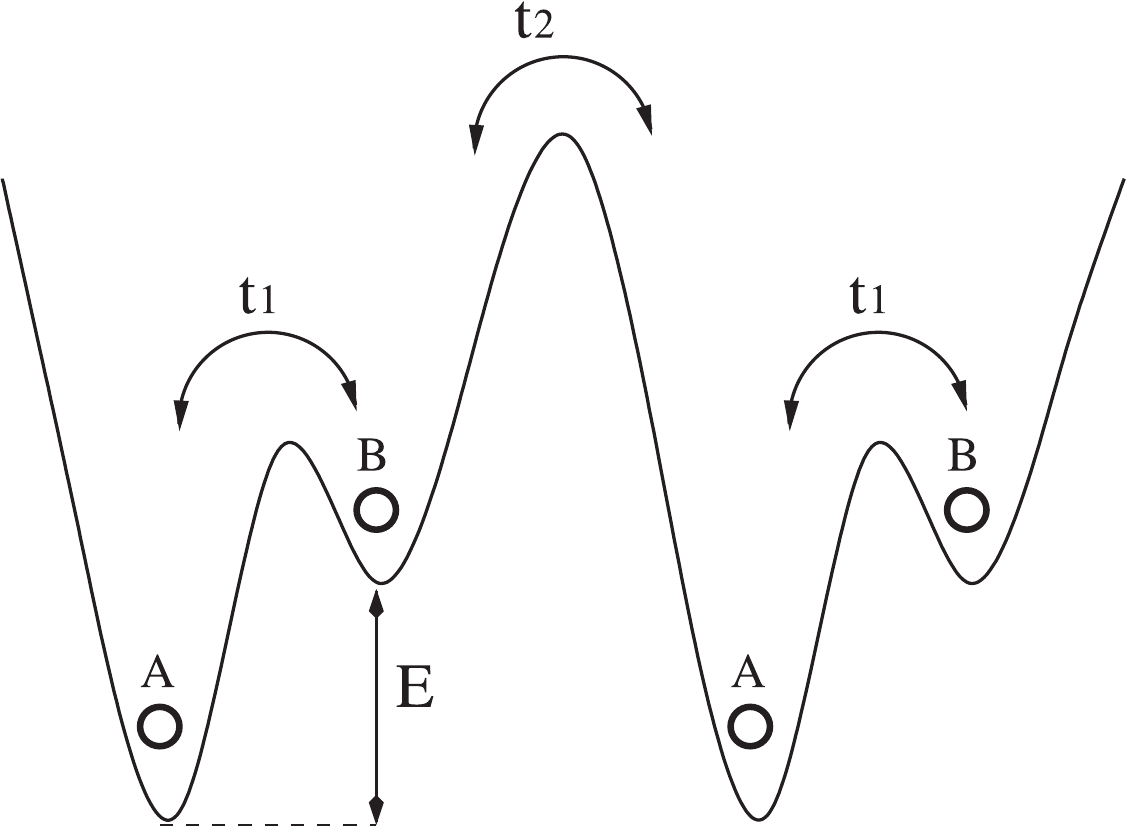}%
\hspace{5mm}
\\%
\parbox[t]{0.45\textwidth}{%
\centerline{(a)}%
}%
\hfill%
\parbox[t]{0.45\textwidth}{%
\centerline{(b)}%
}%
\caption{(a) Schematic picture of a double-well lattice in 3D case, thin lines with arrows correspond to hopping with parameter $t_2$ and thick lines correspond to hopping with parameter $t_1$.
 (b) Double-well potential corresponding to the
 cross  section for the dashed line in (a), $x$ direction.
}
\label{001}
\end{figure}

In this work we consider the thermodynamical
properties of the
BFHM on a lattice with two nonequivalent sublattices.
We consider the hard-core limit (infinite on-site boson-boson interaction).
The geometry of the considered lattice is shown in figure~\ref{001},
where we take into account two parameters of the fermionic hopping ($t_1$ and $t_2$) and bosonic hopping
 ($\Omega_1$ and $\Omega_2$).
 This type of lattices can be experimentally realized.
For example, the lattice with similar topology was  experimentally realized in~\cite{sebby}.

Using  the pseudospin formalism, the Hamiltonian of the model  is written in the following form
 \begin{eqnarray}\label{ham}
 H&=&-\sum_{ij} \Omega_{i_{\textrm A}j_{\textrm B}} [S^+_{i_{\textrm A}} S^-_{j_{\textrm B}}
+S^+_{j_{\textrm B}} S^-_{i_{\textrm A}}]
 - \sum_{ij} t_{{i_{\textrm A}}{j_{\textrm B}}} [c^+_{i_{\textrm A}}
 c_{j_{\textrm B}}+c^+_{j_{\textrm B}} c_{i_{\textrm A}}] \nonumber\\
&& + \sum_{i,\sigma} (g S^z_{i_\sigma} n_{i_\sigma}-\mu_\sigma n_{i_\sigma}
- h_{\sigma }S^z_{i_\sigma}).
 \end{eqnarray}
 The pseudospin variable  $S_i^z=1/2$ when a boson
  is present in a site $i$ (boson concentration $n_{\textrm b}=S^z+1/2$),
$c^+_{i}$ and $c_{i}$ are fermionic  creation and annihilation operators,
      respectively.
The first and the second terms in equation (\ref{ham})) are responsible for
nearest neighbour boson and fermion hopping, respectively;
$g$-term accounts for the boson-fermion interaction energy.
 We consider the grand canonical system
 and  introduce the bosonic and fermionic chemical potentials $h$ and $\mu$,
 respectively, $\sigma=A,B$ is a sublattice index.
 Notice that $h_{\textrm A}=h$, $\mu_{\textrm A}=\mu$,
$h_{\textrm B}=h-E$, $\mu_{\textrm B}=\mu-E$ (see figure~\ref{001}).

 We use the mean field approximation (MFA):
 \begin{eqnarray}
    g n_i S^z_i &\rightarrow& g \langle n_i \rangle S^z_i +
  g n_i \langle  S^z_i \rangle   -
  g \langle  n_i \rangle   \langle  S^z_i \rangle, \nonumber \\
   \Omega S^+_iS^-_j &\rightarrow& \Omega \langle S^+_i \rangle S^-_j +
  \Omega S^+_i \langle  S^-_j \rangle   -
  \Omega \langle  S^+_i \rangle   \langle  S^-_j \rangle,
\end{eqnarray}
 which is appropriate in the case of weak boson-fermion interaction when the fermionic band is not split due to
 this interaction.
Application of the mean field approximation to strongly correlated systems in the limit of a weak on-site correlation makes it possible to satisfactorily describe their properties. For example, application of the similar approximation to the fermionic Hubbard model in the limit of the weak on-site electron correlation makes it possible to describe its magnetic properties.
 In our previous papers (see, for example,~\cite{mysakovych12}), we considered the case $\Omega=0$ (in this case the model is similar to the Falicov-Kimball model) and decoupled the on-site interaction term in a similar way. We  showed that in the case of weak coupling, our results qualitatively agree with those obtained within the framework of the dynamical mean field theory for the Falicov-Kimball model.
In addition, in the case of the Bose-Hubbard model, the kinetic energy term is often considered within the mean field approach.

The Hamiltonian, therefore, is written as follows:
 \begin{eqnarray}
   H &=&- \, 2\Omega\sum_{i} (
 S^x_{i_{\textrm A}} \langle S^x_{\textrm B} \rangle
+
 S^x_{i_{\textrm B}} \langle S^x_{\textrm A} \rangle)
+N\Omega \langle S^x_{\textrm A} \rangle \langle S^x_{\textrm B} \rangle
-\sum_{ ij}
t_{{i_{\textrm A}}{j_{\textrm B}}} [c^+_{i_{\textrm A}} c_{j_{\textrm B}}+c^+_{j_{\textrm B}} c_{i_{\textrm A}}] \nonumber\\
&& + \sum_{i,\sigma} [
( g\langle S^z_{\sigma} \rangle -\mu_\sigma) n_{i_\sigma}
- (h_{\sigma }-g\langle n_{\sigma}\rangle )S^z_{i\sigma}]
-g\frac{N}{2}\langle n_{\textrm A} \rangle\langle S^z_{\textrm A} \rangle
-g\frac{N}{2}\langle n_{\textrm B} \rangle\langle S^z_{\textrm B} \rangle.\nonumber
\end{eqnarray}
 Here, $N$ is the number of lattice sites and $\Omega\equiv\sum_j\Omega_{i_{\textrm A} j_{\textrm B}}$
 (we can see that in the applied approximation
 the difference between $\Omega_1$ and $\Omega_2$ does not play any role).

 At first we diagonalize the pseudospin part of the Hamiltonian using the
 unitary transformation in the pseudospin subspace:
  \begin{eqnarray}
   S^z_{\alpha}&=&\sigma^z_{\alpha}\cos\theta_\alpha+\sigma^x_{\alpha}
   \sin\theta_\alpha \, ,\nonumber\\
    S^x_{\alpha   }&=&\sigma^x_{\alpha  }\cos\theta_\alpha  -\sigma^z_{\alpha  }
   \sin\theta_\alpha \, ,\nonumber\\
   \sin \theta_\alpha &=& -\frac{2\Omega \langle S^x_\beta \rangle}{{\lambda_\alpha}  }\, ,
    \qquad
    \cos\theta_\alpha \ \ = \ \  \frac{h_{\alpha}-gn_\alpha  }{{\lambda_\alpha}}\, ,\nonumber\\
    {\lambda}_{\alpha  }&=&\sqrt{(gn_{\alpha  }-h_\alpha)^2+(2 \Omega \langle S^x_\beta \rangle )^2}, \nonumber\\
H&=&\sum_{{\bf k} \alpha}(g \langle S^z_\alpha\rangle -\mu_\alpha) c^{+}_{{\bf k} \alpha} c_{{\bf k} \alpha}-
\sum_{{\bf k}} [t_{{\bf k}} c^{+}_{{\bf k} {\textrm A}} c_{{\bf k} {\textrm B}}+
t^*_{{\bf k}}c^{+}_{{\bf k} {\textrm B}} c_{{\bf k} {\textrm A}}]
- \sum_{i\alpha} \lambda_\alpha \sigma^z_{i\alpha}
\nonumber\\&-&
\sum_\alpha\frac{N}{2} g\langle S^z_\alpha \rangle
 \langle n_\alpha \rangle
+N\Omega \langle S^x_{\textrm A} \rangle \langle S^x_{\textrm B} \rangle,  \quad  \alpha,\beta=A,B,  \quad \alpha\neq\beta,
      \end{eqnarray}
here we passed to ${{\bf k}}$- representation, $t_{{\bf k}}=|t_{{\bf k}}|\re^{\ri\gamma}$.
In one-dimensional case $|t_{{\bf k}}|^2=t^2_1+t^2_2+2t_1t_2\cos 2{{\bf k}}$, $\cos\gamma=
\frac{(t_1+t_2)\cos{{\bf k}}}{|t_{{\bf k}}|}$,
$\sin\gamma=\frac{(t_1-t_2)\sin{{\bf k}}}{|t_{{\bf k}}|}$.
Similarly, in three-dimensional case for the considered  lattice geometry (see figure~\ref{001})
 $|t_{{\bf k}}|^2=t^2_1+t^2_2+2t_1t_2\cos 2{{k}}_x+4t_2(\cos{{ k}}_y+\cos{{ k}}_z)
 \cos{{ k}}_x(t_1+t_2)+
4t^2_2 (\cos{{ k}}_y+\cos{{ k}}_z)^2 $.
 To diagonalize the fermionic part of the Hamiltonian we perform the following unitary transformation
\begin{eqnarray}
c_{{\bf k} {\textrm A}}&=&   a_{{\bf k} {\textrm A}} \cos{\phi}+
 a_{{\bf k} {\textrm B}} \sin\phi \re^{\ri\gamma} ,\nonumber\\
c_{{\bf k} {\textrm B}}&=& - a_{{\bf k} {\textrm A}} \sin{\phi} \re^{-\ri\gamma}
 + a_{{\bf k} {\textrm B}} \cos\phi     ,\\
 \cos 2\phi&=&\frac{gS^z_{\textrm B}-gS^z_{\textrm A}+E}{2\sqrt{|t_{{\bf k}}|^2
+\left(\frac{gS^z_{\textrm B}-gS^z_{\textrm A}+E}{2}\right)^2}}\,,
 \qquad  \sin 2\phi \ \ = \ \ \frac{-|t_{{\bf k}}|}{\sqrt{|t_{{\bf k}}|^2+\left(\frac{gS^z_{\textrm B}
-gS^z_{\textrm A}+E}{2}\right)^2 }}\,.
      \end{eqnarray}
 It should be noted that fermionic spectrum is always split when $t_1\neq t_2$, this is true
even in the case when
$\langle S^z_{\textrm A}\rangle=\langle S^z_{\textrm B}\rangle$ (when
$\langle S^z_{\textrm A}\rangle\neq\langle S^z_{\textrm B}\rangle$ there is an additional splitting with
 the gap $g|\langle S^z_{\textrm A}\rangle
-\langle S^z_{\textrm B}\rangle|$~\cite{mysakovych1}).
In one-dimensional case, this splitting is of the order of $2|t_1-t_2|$.
In what follows we will discuss the effect of the splitting on the phase transition picture.

 Now we can write
\begin{eqnarray}
  H&=&\sum_{{\bf k} \alpha}\tilde{\lambda}_{{{\bf k}}\alpha}a^+_{{{\bf k}},\alpha}
a_{{{\bf k}},\alpha}
- \sum_{i\alpha} \lambda_\alpha \sigma_{i\alpha}
-
\sum_\alpha\frac{N}{2} g\langle S^z_\alpha \rangle
 \langle n_\alpha \rangle
+N\Omega \langle S^x_{\textrm A} \rangle \langle S^x_{\textrm B} \rangle,\nonumber\\
\tilde{\lambda}_{\bf{k}\alpha}&=&g\frac{\langle S^z_{\textrm A} \rangle+\langle S^z_{\textrm B} \rangle +E}
{2}+(-1)^{\alpha} \sqrt{\left(\frac{ g\langle S^z_{\textrm A} \rangle  - g \langle S^z_{\textrm B} \rangle-E }{2}\right)^2+|t_{\bf{k}}|^{2}}\,.
      \end{eqnarray}

Now we are ready to obtain the system of equations for bosonic and fermionic concentrations
\begin{eqnarray}
\label{n_mod}
\langle
 n_{\alpha}
 \rangle
&{=}&\frac{1}{N}\sum_{{\bf k} }\left\{\frac{1{+}\cos (2 \phi)}{2}
 \left[\exp\left({\frac{\tilde{\lambda}_{{\bf k} \alpha}{-}{\mu}}{T}}\right){+}1\right]^{-1}{+}\frac{1{-}\cos (2 \phi)}{2}
 \left[\exp\left({\frac{\tilde{\lambda}_{{\bf k} \beta}{-}{\mu}}{T}}\right){+}1\right]^{-1}\right\};\nonumber\\
\langle S^z_{\alpha} \rangle&=&\frac{h_{\alpha}-gn_{\alpha}}{2
{\lambda}_{\alpha}}
 \tanh\left(\frac{\beta {\lambda}_{\alpha}}{2}\right); \qquad
 \langle S^x_\alpha \rangle \ \ = \ \ \frac{2\Omega \langle S^x_\beta \rangle  }
{2{\lambda}_\alpha}\tanh\left(\frac{\beta {\lambda}_{\alpha}}{2}\right).
      \end{eqnarray}
The grand canonical potential can be written as~\cite{mysakovych1}
\begin{eqnarray}
\label{Potent_mod}
   \frac{ \Phi}{{N}/{2}}&=&-\frac{T}{N}\sum_{{\bf k}}
      \ln\left\{\left[1+\exp\left({\frac{\mu-\tilde{\lambda}_{{\bf k} {\textrm A}}}{T}}\right)\right]
      \left[1+\exp\left({\frac{\mu-\tilde{\lambda}_{{\bf k}{\textrm B}}}{T}}\right)\right]\right\}\nonumber\\
     && - T\ln\left(4\cosh\frac{\beta{\lambda}_{\textrm A}}{2}
      \cosh\frac{\beta{\lambda}_{\textrm B}}{2}\right)-
g(\langle n_{\textrm A} \rangle \langle S^z_{\textrm A} \rangle
+  \langle  n_{\textrm B}  \rangle \langle S^z_{\textrm B} \rangle )
 +2\Omega \langle S^x_{\textrm A} \rangle \langle S^x_{\textrm B} \rangle.
              \end{eqnarray}

It was shown in~\cite{mysakovych1} that coming from the set of the mean-field equations~(\ref{n_mod})
 we can obtain the condition of the appearance of nonzero values of
 $\delta n=\langle n_{\textrm A} \rangle - \langle n_{\textrm B} \rangle$,
$\delta S^z=\langle S^z_{\textrm A} \rangle - \langle S^z_{\textrm B} \rangle$,
$\delta S^x=\langle S^x_{\textrm A} \rangle - \langle S^x_{\textrm B} \rangle$
(which
play the role of the order parameter for the modulated phase) and this condition coincides with the condition when the static  density-density correlator
$\langle S^z S^z \rangle_{{\bf q}=\pi}(\omega=0)$ calculated in the random phase approximation diverges.

In what follows we  consider a three-dimensional case.
Coming from the set of equations~(\ref{n_mod}),  it can be shown that $\langle S^x \rangle= 0$  when $\Omega<2T$. Therefore, at finite temperature we can consider the transition from the uniform nonsuperfluid  phase (at low temperatures this is Mott insulating (MI) phase) to the charge density wave (CDW) phase for small values of the bosonic hopping parameter ($\Omega<2T$).
We use the equations for averages~(\ref{n_mod}) and the expression for the grand canonical potential~(\ref{Potent_mod}) to find thermodynamically stable states.

\begin{figure}[!b]
\includegraphics[width=0.45\textwidth]{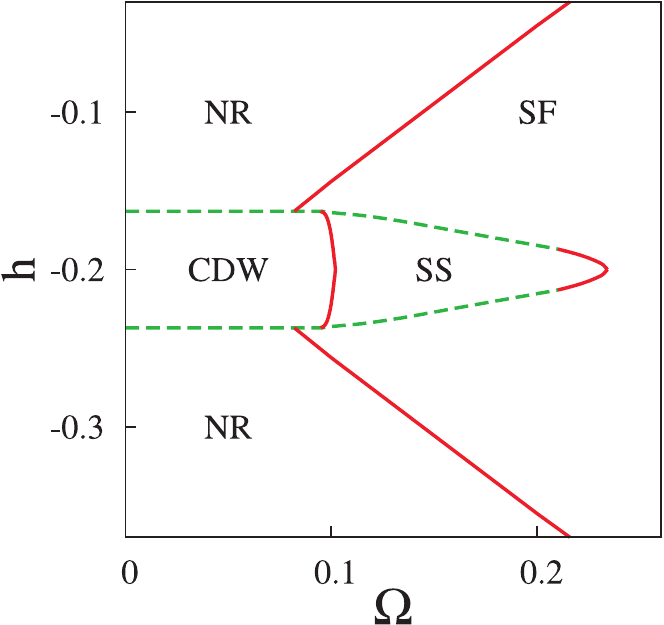}%
\hfill%
\includegraphics[width=0.45\textwidth]{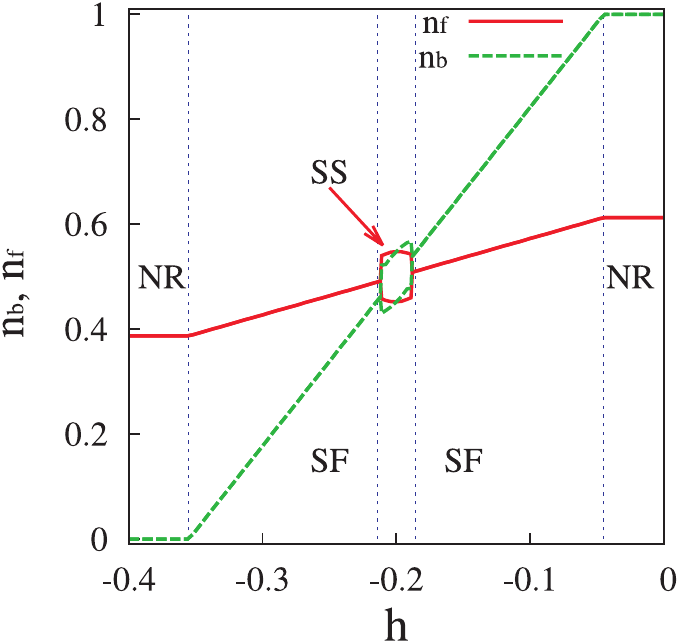}%
\hspace{5mm}
\\%
\parbox[t]{0.45\textwidth}{%
\centerline{(a)}%
}%
\hfill%
\parbox[t]{0.45\textwidth}{%
\centerline{(b)}%
}%
\caption{ (a) Phase diagrams in the ($h-\Omega$) plane for
$t_1=t_2=1/4$, $g=-0.4$, $E=0$,    $\mu=0$, $T=0.005$.
Solid (dashed) lines denote the second (first) order phase transition lines.
(b)~Dependencies of the bosonic and fermionic concentrations
on the chemical potential of  bosons at $\Omega=0.2$.
}
\label{4}
\end{figure}

\begin{figure}[!t]
\includegraphics[width=0.45\textwidth]{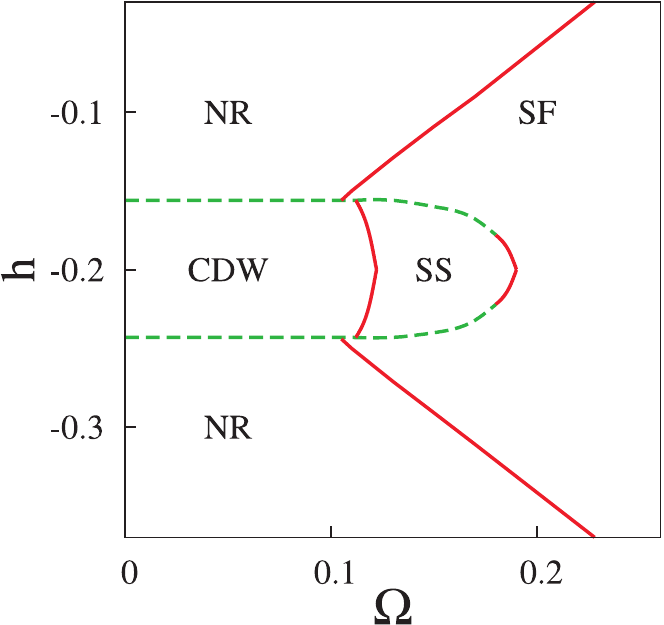}%
\hfill%
\includegraphics[width=0.45\textwidth]{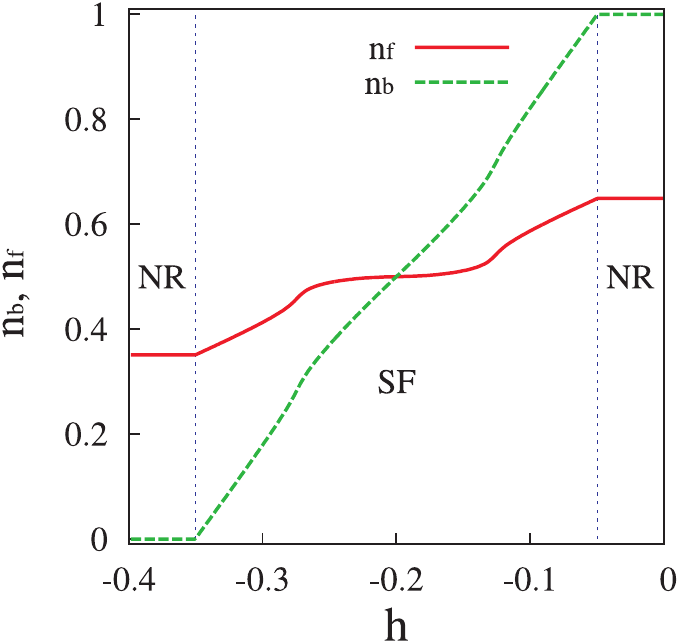}%
\hspace{5mm}
\\%
\parbox[t]{0.45\textwidth}{%
\centerline{(a)}%
}%
\hfill%
\parbox[t]{0.45\textwidth}{%
\centerline{(b)}%
}%
\caption{ (a) Phase diagram in the ($h-\Omega$) plane for
$t_1=1/4,t_2=1/6$, $g=-0.4$,   $E=0$,  $\mu=0$, $T=0.005$.
Solid (dashed) lines denote the second (first) order phase transition lines.
(b)~Dependencies of the bosonic and fermionic concentrations
on the chemical potential of  bosons at $\Omega=0.21$.
}
\label{5}
\end{figure}

\begin{figure}[!b]
\includegraphics[width=0.45\textwidth]{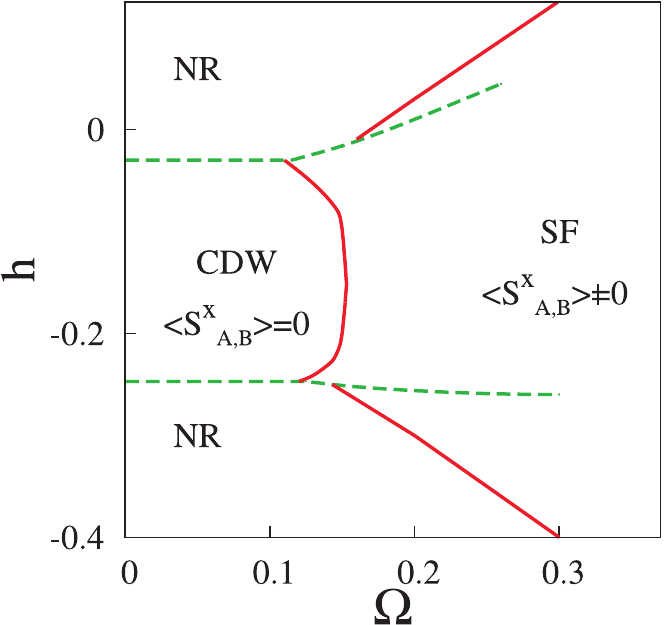}%
\hfill%
\includegraphics[width=0.43\textwidth]{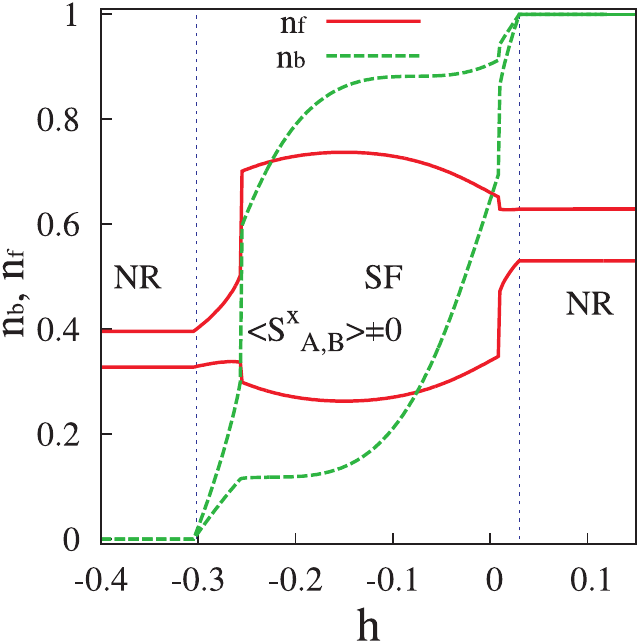}%
\hspace{5mm}
\\%
\parbox[t]{0.45\textwidth}{%
\centerline{(a)}%
}%
\hfill%
\parbox[t]{0.45\textwidth}{%
\centerline{(b)}%
}%
\caption{ (a) Phase diagram in the ($h-\Omega$) plane for
$t_1=1/4,t_2=1/4$, $g=-0.4$,   $E=0.1$,  $\mu=0$, $T=0.005$.
Solid (dashed) lines denote the second (first) order phase transition lines.
(b)~Dependencies of the bosonic and fermionic concentrations
on the chemical potential of  bosons at $\Omega=0.2$.
}
\label{6}
\end{figure}

In figures~\ref{4}, \ref{5}, \ref{6} we show
 the phase diagrams in the plane ($h-\Omega$) and the
 dependencies of the bosonic and fermionic concentrations on bosonic chemical potential at low temperature
 for different cases when $t_1=t_2$, $t_1\neq t_2$, $E=0$, $E\neq0$.
 We consider the regime of the fixed fermionic
chemical potential and from figures~\ref{4}, \ref{5}, \ref{6}
 we can see that
the phase transition from the uniform  to chess-board phase can be of the second (solid line)
 or  first (dashed line) order.
The existence of the phase transition of the first order leads to phase separation (in the regime of the fixed concentrations) into the uniform and CDW phases (this was illustrated in~\cite{mysakovych2}).
 The presence of anisotropic hopping ($t_1\neq t_2$) leads to the narrowing of the region of the supersolid (SS) phase (this phase is characterized by the simultaneous presence of a density wave and phase order in the condensate).
 This is due to the appearance of the  above mentioned gap in
the fermionic spectrum leading to the plato-like behaviour
 in the dependence
 of the fermionic concentration
 on the bosonic chemical potential in the SF phase,
see figure~\ref{5}~(b).
In our numerical calculations we considered the case $t_1=1/4$, $t_2=1/6$, but similar conclusion (about the narrowing of the SS region) is also valid for other values of $t_1$, $t_2$ (for example, we also calculated
 the phase
 diagrams for the cases  $t_1=1/3$, $t_2=1/4$, and $t_1=1/6$, $t_2=1/4$, but here we do not present these diagrams because they are similar to that shown in figure~\ref{5}). With an increase of  temperature,  the first order phase transition transforms into the second one and then disappears at some critical value of the temperature, as it was discussed in our previous work~\cite{mysakovych1} at the construction of $T-h$ diagrams in the case $t_1=t_2$.

\begin{wrapfigure}{i}{0.5\textwidth}
\centerline{\includegraphics[width=65mm]{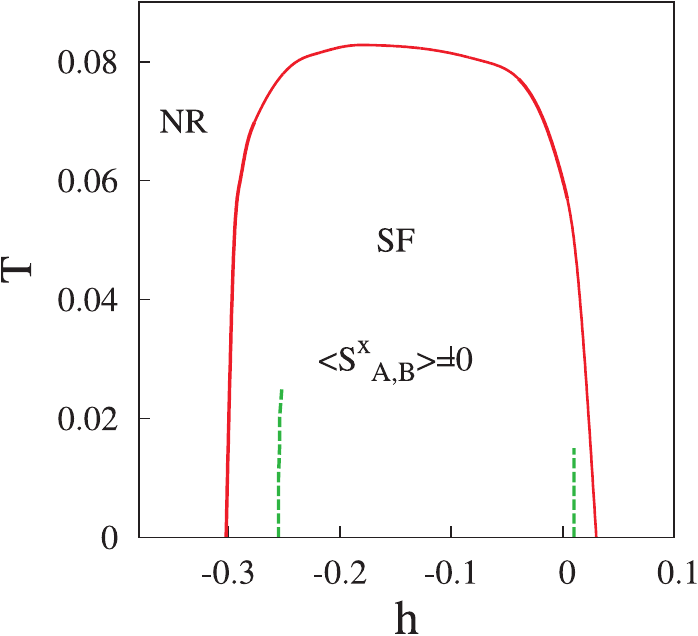}}
\caption{ Phase diagram in the ($T-h$) plane for
$t_1=1/4,t_2=1/4$, $g=-0.4$,   $E=0.1$,  $\mu=0$, $\Omega=0.2$.
Solid (dashed) lines denote the second (first) order phase transition lines.}
\label{7}
\end{wrapfigure}

 The case $E\neq 0 $ is illustrated in figure~\ref{6}. As we can see, the presence of the modulated potential leads to the  modulation of bosonic and fermionic concentrations in sublattices (at any finite value of $E$ we always have such a modulation),
 we call this phase a normal (NR) nonsuperfluid phase
to distinguish it from a true
charge-density-wave  phase, for which the translational invariance
is spontaneously broken  due to the presence
of boson-fermion interaction. In our approach,
we can distinguish between these two phases (NR and CDW) only when the phase transition between them is of the first order with jumps
of the fermionic and bosonic concentrations.
As we can see, the phase diagram is asymmetric in the case $E\neq 0$ in comparison with the case $E=0$ (see figures~\ref{4},~\ref{5}) because the chemical potential of particles
on one of the sublattices is shifted away from the half-filling case.
When we increase the value of the double-well potential $E$, the line of the first order phase transition
between modulated phases at higher values of the bosonic chemical potential ($h\approx-0.03$ at $\Omega=0$ in figure~\ref{6}~(a))
gets shorter  and the line of the first order phase transition at lower values of the bosonic chemical potential
($h\approx-0.25$ at $\Omega=0$ in figure~\ref{6}~(a)) gets longer (here we do not present phase diagrams for other values of $E$ because  this tendency is also seen in figure~{\ref{6}~(a)}).

The presence of the first order phase transition is connected with the effective interaction between bosons via fermions. As the temperature increases, this effective interaction decreases and at some critical value of temperature this first order phase transition disappears (see figure~\ref{7}).
At low temperature the bosonic concentration is almost
the same in both sublattices ($n_{\textrm b}=0$ or $n_{\textrm b}=1$ depending on the value of the bosonic chemical potential) when we go into NR phase (at zero temperature in the case of the Bose-Hubbard model this phase is sometimes referred to as a hole vacuum or particle vacuum~\cite{iskin} when $n_{\textrm b}=1$ or $n_{\textrm b}=0$, respectively),
see figure~\ref{6}~(b). We do not use the notation ``SS phase'' in figure~\ref{6} because we cannot distinguish between SS and SF phase (for the case of a true SS phase, the translational invariance is spontaneously broken rather than
due to the presence of the modulated potential).

 \section{Conclusions}

The Bose-Fermi-Hubbard model
on a lattice with two nonequivalent sublattices has been  investigated in this work,
and this type of optical lattices can be realized experimentally (see, for example,~\cite{sebby}).
We  considered the hard-core boson limit and investigated the thermodynamics of the model for the
case of weak boson-fermion interaction.
The phase diagrams in the plane of the chemical potential of bosons-bosonic hopping parameter were built and
it was shown that
the transition from the uniform  to modulated phase can be of the first or the second order
(in the regime of the fixed concentrations this leads to phase separation).
The  gap in the fermionic spectrum appears due   to
the presence of anisotropic hopping  and
 this  causes the
narrowing of the
 parameter region of supersolid phase existence. The presence of the effective interaction between bosons via fermions leads to the first order phase transition between modulated phases at the presence of the double-well potential $E$.
It should be noted that we considered thermodynamics in the mean field approximation
 and our future investigations will be devoted to the
corrections beyond
the mean field approximation.

\newpage

\ukrainianpart

\title{
Модель Бозе-Фермі-Габбарда на ґратці із двома нееквівалентними підґратками}

\author{
Т.С. Мисакович}

\address{
Інститут фізики конденсованих систем НАН України,
вул. І.~Свєнціцького,~1, 79011~Львів, Україна}

\makeukrtitle

\begin{abstract}

У цій роботі досліджено фазові переходи у системах, що описуються моделлю Бозе-Фермі-Габбарда на ґратці
 з двома нееквівалентними підґратками.
 Розглянуто випадок жорстких бозонів та використано псевдоспіновий формалізм.
 Побудовано фазові діаграми у площині хімічний потенціал бозонів-бозонний параметр перескоку.
 Показано, що у випадку анізотропного перескоку область існування так званої ``суперсолід'' фази  звужується.

\keywords
 модель Бозе-Фермі-Габбарда, оптичні ґратки, фазові переходи

\end{abstract}

\end{document}